\documentclass[graybox]{svmult}
\def\email#1{\emailname: \href{mailto:#1}{#1}}

\usepackage[
  textwidth=117mm,
  centering
]{geometry}

\usepackage{type1cm}          
\usepackage{makeidx}          
\usepackage{graphicx}         
\usepackage{multicol}         
\usepackage[bottom]{footmisc} 
\usepackage{amssymb,amsmath,dsfont,amsthm}
\usepackage{soul,multirow}
\usepackage{graphicx}
\usepackage{subcaption}
\usepackage[numbers]{natbib}
\usepackage[linesnumbered, ruled, lined, boxed]{algorithm2e}
\usepackage{booktabs,array}
\usepackage{newtxtext}
\usepackage[varvw]{newtxmath}       
\usepackage{url}
\usepackage{hyperref}
\hypersetup{
  colorlinks,
  linkcolor=Blue4,
  citecolor=Blue4,
  urlcolor=Blue4}

\newcommand \one  {\mathds{1}}  
\newcommand \Rset {\mathbb{R}}  
\newcommand \Sset {\mathbb{S}}  

\newcommand \Jcal {\mathcal{J}} 

\newcommand \PF       {\pi_F}                       
\newcommand \hatP     {{\hat\pi}}                   
\newcommand \hatPF    {\hatP_F}                 
\newcommand \hatPFref {\hatP_{F,\mathrm{ref}}}  

\newcommand \btheta {\boldsymbol{\theta}}  
\newcommand \bTheta {\boldsymbol{\Theta}}  
\newcommand \dtheta {\mathrm{d}\btheta}    
\newcommand \modF   {F'}                   
\newcommand \Prob   {\mathsf{P} }          

\begin{document}

\title*{%
  Directional subset simulation method for reliability analysis%
}%
\author{%
  Oindrila Kanjilal\orcidID{0000-0003-1855-7483} %
  and Julien Bect\orcidID{0000-0002-0867-0215}%
}%
\institute{%
  Oindrila Kanjilal \at Georg Nemetschek Institute, Technical University of Munich\\
  80333 Munich, Germany. \email{oindrila.kanjilal@tum.de}%
  \and Julien Bect \at Université Paris-Saclay, CNRS, CentraleSupélec, Laboratoire des signaux et systèmes\\
  91190 Gif-sur-Yvette, France. \email{julien.bect@centralesupelec.fr}%
}%

\maketitle

\abstract{%
  Estimating the probabilities of rare failure events is a key
  challenge in the reliability analysis of physical systems. Subset
  simulation (SS) is a very popular adaptive Monte Carlo method for
  this problem. In SS, the small failure probability is evaluated as a
  product of larger conditional probabilities by iteratively sampling
  a sequence of nested sub-domains of the parameter space,
  encompassing the target failure domain of interest, using Markov
  chain Monte Carlo methods. For failure domains with multiple modes,
  the Markov chain samples used to explore the intermediate levels of
  SS can be trapped in a confined region of the input parameter space,
  leading to inaccurate failure probability estimates. In this
  contribution, we propose the directional subset simulation (dSS)
  method for this problem, which uses concepts from directional
  sampling to informedly propagate samples towards failure. This is
  accomplished through a novel selection of the intermediate failure
  domains, which preserves samples in several directions in the
  parameter space in each intermediate level. The merits of the dSS
  method are illustrated through a selection of numerical examples.%
}

\section{Introduction}
\label{sec:1}

One of the main goals of the condition assessment of engineering
systems is to determine the safety level, or reliability, by
accounting for the impact of uncertainties on system performance. %
These uncertainties arise from various sources, including loading,
material properties, geometry and deterioration parameters. %
Computational reliability analysis focuses on evaluating the
probability of failure:
\begin{equation}
  \PF
  \;=\; \Prob\left( \bTheta \in F \right) 
  \;=\; \Prob\left( g\left( \bTheta \right) \le 0 \right)
  \;=\; \int_\Sset \one_F(\btheta)\, p(\btheta)\, \dtheta,
  \label{rel_int}
\end{equation}
where $\bTheta$ is an $n$-dimensional random vector of
uncertain quantities, %
taking values in some set $\Sset \subseteq \Rset^n$, %
$p$ its probability density function (PDF),
$g:\Sset \to \Rset$ the limit-state function (LSF) and
$F = \left\{ \btheta \in \Sset \mid g(\btheta) \le 0 \right\}$ the
failure domain. %
To simplify notations, we will identify in this article a
subset~$A \subset \Sset$ with the
event~$\left\{ \bTheta \in A \right\}$, writing for
instance~$\PF = \Prob(F)$ instead
of~$\PF = \Prob\left( \bTheta \in F \right)$ as in~\eqref{rel_int}.%

In practical applications, the failure domain may have a complex
geometry, making the evaluation of the reliability integral
challenging. %
A prominent challenge arises when the truncated PDF
$p(\btheta|F) \propto p(\btheta)\one_F(\btheta)$ has multiple dominant
regions, or \emph{modes}, in the parameter space~$\Sset$. %
This situation occurs in system reliability analysis
problems, where failure can be due to a single component of the system 
subjected to multiple failure mechanisms (e.g.,
stress and displacement in a component), or due to simultaneous failure of multiple
components. %
To understand failure behavior in such settings, it is important not
only to compute the global failure probability~$\PF$, but also to
estimate the relative contribution of each mode to the failure event. %
Effective reliability analysis therefore requires methods that
adequately explore all failure modes, which is difficult when
the modes are separated by regions of low probability density.

A simple but versatile method for solving reliability problems with
multi-modal failure domains is Monte Carlo simulation (MCS), which
uses an independent and identically distributed (i.i.d) sample drawn
from~$p$:
\begin{equation}
  \hatPF^{\mathrm{MCS}} = \frac{1}{N}\sum_{i=1}^{N}\one_F\left({\btheta}^{(i)}\right),
  \qquad {\btheta}^{(i)}\overset{\text{i.i.d}}{\sim}{p(\btheta)}.
  \label{brute-force}
\end{equation} 
This method yields an unbiased estimator of~$\PF$, with coefficient of
variation (CoV) $\delta_{\text{MCS}} \approx 1 / \sqrt{N\PF}$. %
Its main drawback is the large sample size~$N$ that is required to
explore the failure domain when the failure probability is small.

Analytical methods based on the FORM/SORM principle (see, e.g.,
\cite{Breitung_PEM_2015} and references therein) do not handle
``natively'' problems with multiple failure mode---since they rely on
an approximation of the LSF around the most probable failure
point---but can be extended to work with several design points
\citep{Kiureghian_SS_1998,Zhang_PEM_2025}.
A widely adopted approach for
estimating small failure probabilities is to use advanced Monte Carlo methods. %
These methods modify the brute-force procedure with specialized
techniques to sample from the important regions of a rare failure
domain in a probabilistically correct and efficient way. %
Several advanced simulations methods have been proposed over the past decades,
which can be broadly grouped as importance sampling (IS) methods 
\citep{Kurtz_SS_2013,Papaioannou_SS_2016,Xian_SS_2024}, subset
simulation \citep{Au_PEM_2001} and its variants \citep{Kanjilal_PEM_2015,Kinnear_PEM_2025},
line sampling methods \citep{Pradlwarter_SS_2007,Papaioannou_SS_2021}
and directional simulation \citep{Ditlevsen_CS_1990,Nie_PEM_2004,Zhang_RESS_2022},
among others. %
Discussions on the strengths and limitations of the different
approaches for reliability analysis as well as insight on the relative
performance is available in \citep{Tabandeh_SS_2022,Song_PEM_2023}.

In this paper, we focus on the widely used subset simulation (SS)
method \citep{Au_PEM_2001}. %
The key idea in SS is to introduce a nested sequence of intermediate
failure domains, or \emph{subsets}, such that
the small failure probability can be decomposed into a product
of larger conditional probabilities. %
The probabilities are estimated by sampling from the conditional PDFs 
using Markov chain Monte Carlo (MCMC) \citep{MCMC_in_practice}. %
SS preserves the versatility of brute-force MCS as it does not require
any prior information about the system behavior and is universally
applicable to all system types. %
However, it is well known that for certain geometries of the LSF
\citep{Breitung_RESS_2019}, e.g., when there are sudden sharp changes
in the LSF values or discontinuity in the failure domain, the Markov
chains in an intermediate subset can get trapped in a confined region
of the parameter space causing the modes of the subsequent conditional
PDFs to remain unexplored, see Fig. \ref{Ex4p1_c1_samples}a for an illustration. %
This can cause several important regions of the target failure domain to go 
undetected during reliability analysis, resulting in
underestimation of the failure probability and (or) high sampling
variance. Various enhancements of the original SS method have been proposed to
overcome this weakness \citep{Sharma_PEM_2023,Rakshi_MSSP_2021, Kinnear_PEM_2025}. %

In this paper, we propose a new framework for subset simulation, which
we name directional subset simulation (dSS). %
In dSS, the key idea is to direct the Markov samples towards the
important failure regions by partitioning the parameter space into a
set of communicating bins, and individually exploring each bin with
suitable subsets. %
The subsets define a set of local intermediate failure domains in each
bin and are constructed such that the global conditional failure
probability over the parameter space adheres to an prescribed large
value. %
MCS is used to allocate an initial population of samples in each bin,
which are propagated to the failure domain by sequentially sampling
the subsets using global MCMC moves. %

The paper is organized as follows. %
In the next section, we briefly summarize the conventional SS method. %
This is followed by the description of the proposed dSS method in Section \ref{sec:3}. %
In Section \ref{sec:numerical-examples} we demonstrate the performance of the dSS method on two
numerical examples. %
Finally, we present the conclusions in Section \ref{sec:conclusions} and provide
directions for future research.

\section{Subset simulation method}
\label{sec:2}

Consider a failure event
$F=\{\btheta \in \mathbb{S} \mid g(\btheta) \leq 0\}$. %
Let $\infty=\gamma_0>\gamma_1>\gamma_2 > \cdots > \gamma_T = 0$ be a
decreasing sequence of thresholds, which defines a decreasing sequence
\begin{equation}
  \label{equ:subsim:decreasing-seq}
  \Sset = F_0
  \;\supset\; F_1
  \;\supset\; F_2
  \;\supset\; \cdots
  \;\supset\; F_T = F
\end{equation}
of subsets, where
$F_t = \{\btheta \in \mathbb{S} \mid g(\btheta) \leq \gamma_t\}$,
$0 \le t \le T$. Since $F_{t} \subset F_{t-1}$, we have
\begin{equation}
  \Prob(F_{t}) \,=\, \Prob(F_{t}|F_{t-1})\, \Prob(F_{t-1}),
  \quad 1 \le t \le T.
\end{equation}
Since $\Prob(F_0) = 1$, the target failure probability can be
expressed as $\PF = \prod_{t=1}^Tp_t$, with
$p_t := \Prob(F_{t}|F_{t-1}) =
\int_\Sset\one_{F_t}(\btheta)f_{t-1}(\btheta) \dtheta$, where
$f_{t-1}$~is the truncated density
\begin{equation}
  \label{equ:SS:truncated-density}
  f_{t-1}(\btheta)
  = \frac{%
    \one_{F_{t-1}}(\btheta)\, p(\btheta)%
  }{%
    \int_{\btheta \in \mathbb{S}}\one_{F_{t-1}}(\btheta)\, p(\btheta)\, \dtheta%
  }.
\end{equation}
The central idea in SS is then to estimate~$\PF$ by estimating the
conditional failure probabilities~$p_t$, $1 \le t \le T$, expecting
efficiency gain when these probabilities are not small. %
There are two key challenges in implementing this idea.

First, estimating the conditional probabilities by simulation requires
the efficient generation of samples from the truncated
densities~$f_{t}$, $1 \le t \le T-1$, which in general is not
trivial. %
This issue is resolved by using Markov chain Monte Carlo
\cite{Papaioannou_PEM_2015}, in a sequential Monte Carlo
\cite{Cerou_StatComput_2012} framework: %
the samples that belong to the failure domain~$F_{t}$ at level~$t$
form the seeds for MCMC sampling at the next level. %

Second, the intermediate thresholds $\gamma_t$, $1 \le t \le T-1$,
must be selected such that the $p_t$s are neither too small (to avoid
ending up with a rare event estimation problem at each subset) nor too
large (otherwise the method would require a large value of~$T$). %
To achieve this goal, the intermediate thresholds are chosen
adaptively, in such a way that estimated conditional failure
probabilities~$p_t$, $1 \le t \le T-1$, are all equal to a given value
$\rho \in [0.1, 0.3]$ \citep{Zuev_CS_2012}. %
In other words, $\gamma_t$ is the quantile of order~$\rho$ of the LSF
values $g\big(\btheta_{t-1}^{(i)}\small)$, where
$\btheta_{t-1}^{(i)}$, $1 \le i \le N$, are the samples generated at
level~$t-1$.
The procedure is repeated until the estimated $\rho$-percentile
becomes negative, %
at which stage we have $\gamma_T = 0$ and~$F_T = F$. %
An estimate of the probability of failure is then obtained as
\begin{equation}
  \hatPF^{\mathrm{SS}} = \rho^{T-1}\hat{p}_T,
  \label{cSS_estimator}
\end{equation}
where
$\hat{p}_T =
\frac{1}{N}\sum_{i=1}^N\one_F\big(\btheta_{T-1}^{(i)}\big)$. %
The statistical properties of the estimator $\hatPF^{\mathrm{SS}}$
have been discussed in \citep{Au_PEM_2001,Cerou_StatComput_2012}.

\section{Directional subset simulation}
\label{sec:3}

\noindent\textbf{Main idea: more flexible intermediate failure sets}. %
Assume without loss of generality that $\Sset = \Rset^n$ and
$\boldsymbol{\Theta} = (\Theta_1,\cdots,\Theta_n)$ is an
$n$-dimensional standard Gaussian random vector, i.e.,
$p(\btheta) = \prod_{i=}^n\phi({\theta_i})$, where $\phi$~is the PDF
of a standard Gaussian random variable. %
(When the parameters are not Gaussian or not independent, $\btheta$ can
be turned into a standard Gaussian random vector by means of a
suitable iso-probabilistic transformation \citep{StrucRel_AnalysisPrediction}.) %

Let $\left( B_j \right)_{1 \le j \le J}$ denote a partition of the
parameter space~$\Sset$, where each subset~$B_j$, hereafter called
a~\emph{bin}, corresponds to a set of half-lines emanating from the
origin and stretching outwards to infinity (i.e., a linear cone
in~$\Rset^n$). %
In other words, each bin correspond to a set of \emph{directions}.  We
consider for now that the bins are preselected before reliability
estimation. %
We will propose a modified subset simulation algorithm, called
directional subset simulation (dSS), which will ensure that each bin
is properly explored.

Consider a particular bin~$B_j$. %
For a decreasing sequence of threshold levels
$\infty=\gamma_{0,j}>\gamma_{1,j} \geq \gamma_{2,j} \geq \cdots$, the
corresponding sub-domains
\begin{equation} 
  \Gamma_{t,j} = \{\btheta \in B_j \mid g(\btheta) \leq \gamma_{t,j}\}.
\end{equation}
form a nested sequence of subsets of~$B_j$, where $\Gamma_{0,j} = B_j$
and
$\Gamma_{0,j} \supset \Gamma_{1,j} \supset \Gamma_{2,j} \supset
\cdots$. %
In the proposed dSS method, we define intermediate failure domains by
\begin{equation}
  F_t  = \{\btheta \in \mathbb{S} \mid g(\btheta) \leq \gamma_{t}(\btheta)\}, 
\end{equation}
where the $\btheta$-dependent thresholds~$\gamma_{t}$ are bin-wise
constant:
$\gamma_t(\btheta) = \sum_{j=1}^J \gamma_{t,j}\one_{B_j}(\btheta)$. %
The sets $\Gamma_{t,j},1\leq j\leq J$ form a partition of~$F_t$, and
the intermediate failure domains form a decreasing sequence:
$\Sset = F_0 \supset \cdots \supset F_t \supset F_{t+1} \supset
\cdots$ as in~\eqref{equ:subsim:decreasing-seq}. %

It can be readily seen that SS is a special case of the proposed dSS
algorithm, where there is only one bin. %
The proposed framework introduces more flexibility in the choice of
the intermediate failure domains: %
the inclusion of multiple thresholds at each level will make it
possible to ensure, through a careful choice of the local
thresholds~$\gamma_{t,j}$, that each bin remains populated (with high
probability), and is thus properly explored.

\bigbreak

\noindent\textbf{Entering the failure domain}. %
In each bin the limit state~$\{ g = 0 \}$ will be reached at a
different level, if it is reached at all: %
we denote by~$T_j$ the last level~$t$ such that~$\gamma_{t,j} > 0$,
with $T_j = \infty$ if $\gamma_{t,j} > 0$ for all~$t$, and we set
$\gamma_{t,j} = 0$ for all~$t > T_j$. %
The sequence of local thresholds in the bin~$B_j$ thus has the form
(assuming $T_j < +\infty$):
\begin{equation*}
  \infty = \gamma_{0,j}
  \;>\; \gamma_{1,j} \;>\; \cdots \;>\; \gamma_{T_j,j}
  \;> \;\gamma_{T_j+1, j} \;=\; \gamma_{T_j+2,j} \;=\; \cdots \;=\; 0.
\end{equation*}

Let $\Jcal_t \subseteq \{1, \ldots, J\}$ be the set of all bin indices
such that~$\gamma_{t,j} = 0$. %
When $\Jcal_t \neq \varnothing$, i.e., when the limit state has been
reached in some bins, the dSS algorithm stops exploring these bins and
focuses on the remaining bins~$B_j$, $j \not\in \Jcal_t$. %
Accordingly, we define the following intermediate sampling regions:
\begin{equation*}
  \modF_t \;=\; \cup_{j \not\in \Jcal_t}\, \Gamma_{t,j},
\end{equation*}
which form again a decreasing sequence:
$\Sset = \modF_0 \supset \cdots \supset \modF_t \supset
\modF_{t+1} \supset \cdots$. %
Note however that $\modF_t \not\supseteq F$ since the bins
where~$\gamma_{t,j} = 0$ have been removed.

\bigbreak

\noindent\textbf{Sampling from the truncated densities}. %
The dSS algorithm constructs at each level~$t$ an $N$-sample of
(correlated) particles (approximately) distributed according to the
truncated densities $f_t \propto \one_{\modF_t} \, p$, using the
particles from level~$t-1$ in a sequential Monte Carlo framework
exactly as in~SS (see Section~\ref{sec:2}). %

Let $M_t$ denote the number of particles from level~$t - 1$
that belong to $\modF_t$. %
Classical SS constructs an $N$-sample at stage~$t$ using
$\rho^{-1}$~steps of an MCMC chain started from each of these
particles (called \emph{seeds}), which is possible because
$M_t = N \rho$ by construction, and $\rho^{-1}$ is an
integer. %
In dSS, as a consequence of the selection of local thresholds (see
below), $M_t$ will typically be close, but not equal, to
$N\rho$. %
To restore a population of size~$N$, we use residual resampling
\cite{douc2005comparison} to determine the (random) number of particles
produced by MCMC from each seed.
(Note that, in the situation where $M_t = N \rho$ and~$\rho^1$ is an
integer, residual resampling is equivalent to the original subset
simulation approach, which is to take $\rho^{-1}$ steps from each
seed.)

\bigbreak

\noindent\textbf{Selection of the sequence of thresholds}. %
We propose to select the local thresholds~$\gamma_{t,j}$ in a manner
analogous to SS. %
More precisely, let $N_{t, j}$ be the number of particles in bin~$B_j$
at level~$t$, $j \not\in \Jcal_t$, and let
$\btheta_{t,j}^{(i)} \in \Gamma_{t,j}$, $1 \leq i \leq N_{t,j}$, be
the corresponding particles. %
These particles are (approximately) distributed according to the
truncated density~$f_{t,j} \propto \one_{\Gamma_{t,j}}\, p$. %
Then, the next local threshold~$\gamma_{t+1,j}$ in bin~$B_j$ is taken
equal to an estimate of the quantile of order~$\rho$ of the LSF values
$g\bigl( \btheta_{t, j}^{(i)} \bigr)$, $1 \leq i \leq N_{t, j}$, if
this estimate is positive, and to zero otherwise. %
(Since the number of samples $N_{t, j}$ in each bin is not necessarily
an integer multiple of $\rho^{-1}$ as in~SS, %
we use a quantile estimator based on an interpolation of the empirical
cumulative distribution function.)

\bigbreak

\noindent\textbf{Probability estimation and stopping condition.} %
Our goal is to estimate the probability of failure
$\pi_j = \Prob\left( F \cap B_j \right)$ in each bin~$B_j$, and the
total probability of failure~$\PF = \Prob(F) = \sum_{j=1}^J
\pi_j$. %
At level~$t$, we have $\gamma_{t+1, j} = 0$ for $j \in \Jcal_{t+1}$, %
and therefore $T_j \le t$. %
Thus, the probability of failure~$\pi_j$ can be decomposed for these bins as
\begin{equation*}
  \pi_j
  \;=\; \Prob(B_j)\, \textstyle\prod_{t=0}^{T_j} \Prob\left( \Gamma_{t+1, j} \mid \Gamma_{t, j} \right)
  \;=\; \textstyle\prod_{t=0}^{T_j+1} p_{t,j},
\end{equation*}
where
$p_{0,j} = \Prob(B_j) = \int_{\btheta \in
  \Sset}\one_{B_j}(\btheta)p(\btheta)\, \dtheta$ and %
$p_{t, j} = \Prob\left(\Gamma_{t, j}\mid \Gamma_{t-1, j}\right)$ for
$t \geq 1$, and we have the estimate
\begin{equation}
  \hatP_j
  \;=\; \textstyle\prod_{t=0}^{T_j + 1} \hat{p}_{t,j}
  \;=\; p_{0,j}\, \rho^{T_j}\, \hat{p}_{T_j+1, j},
\end{equation}
where $\hat{p}_{t,j} = \rho$ for $1 \le t \le T_j$, and
${\hat{p}}_{T_j+1,j} = \frac{1}{N_{t,j}}
\sum_{i=1}^{N_{t,j}}\one_{F}\Bigl( \btheta_{t,j}^{(i)} \Bigr)$. %
The bin probabilities $p_{0,j}$, $1 \le j \le J$, are assumed to be
known---they will be equal to~$1/J$ in our numerical experiments.

For the remaining bins ($j \not\in \Jcal_{t+1}$ at level~$t$), we are
not yet in a position to estimate the probability~$\pi_j$ accurately,
but we can already estimate an upper-bound, using the fact
that~$T_j \ge t+1$:
\begin{equation*}
  \hatP_{t, j}^{+}
  \;=\; \textstyle\prod_{t=0}^{t + 1}
  \;=\; p_{0,j}\, \rho^{t+1}.
\end{equation*}
We stop the algorithm when
$\sum_{j \not\in \Jcal_{t+1}} \hatP^+_{t,j} \le
\varepsilon_{\mathrm{tol}} \sum_{j \in \Jcal_{t+1}} \hatP_j$, and
define our final estimate as
$\hatPF = \sum_{j \in \Jcal_{t+1}} \hatP_j$. %
(We take $\varepsilon_{\mathrm{tol}} = 0.001$ in the numerical
experiments.)

\begin{remark}
  Note that such a stopping criterion is possible because all the bins
  are explored simultaneously, rather than sequentially, as would be
  the case in a straightforward combination of subset simulation with
  stratified sampling.
\end{remark}

\section{Numerical examples}
\label{sec:numerical-examples}

\subsection{Piecewise linear function}
\label{sec:piecewise-linear}

\begin{figure}[htb]
  \sidecaption
  \includegraphics[width=0.65\textwidth,clip=true,trim=75mm 25mm 100mm 30mm]{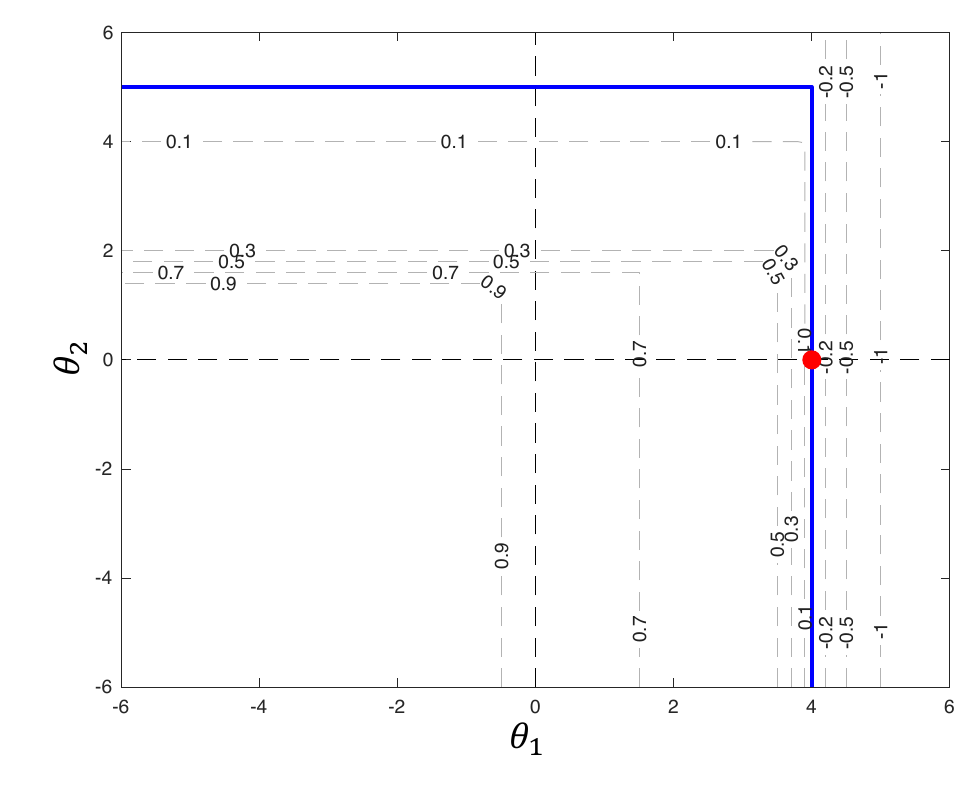}
  \caption{Contour plot of the piecewise linear function in Example
    4.1. The blue line denotes the boundary of the failure domain. The
    red dot is the global design point $(\theta_1 = 4,\theta_2 = 0)$.}
  \label{Ex1Fig1}
\end{figure}

We first consider the problem mentioned in Section~\ref{sec:2}. %
The LSF is given by
\begin{equation}
	g(\theta_1,\theta_2) = \mathrm{min}(g_1,g_2)
\label{Ex1_1}
\end{equation}
where
\begin{equation}
  \begin{aligned}
    g_1(\theta_1,\theta_2)
    &= \left\{
      \begin{array}{ll}
        4 - \theta_1 &\text{for } \theta_1 > 3.5, \\
        0.85 - 0.1\theta_1 &\text{for } \theta_1 \leq 3.5,
      \end{array}\right.
    \\
    g_2(\theta_1,\theta_2)
    &= \left\{
      \begin{array}{ll}
        0.5 - 0.1\theta_2 &\text{for } \theta_2 > 2, \\
        2.3 - \theta_2 &\text{for } \theta_2 \leq 2.
      \end{array}\right.
  \end{aligned}
  \label{Ex1_2}
\end{equation}
Here $\theta_1$ and $\theta_2$ are both standard Gaussian random
variables. %
Fig.~\ref{Ex1Fig1} depicts the contour plot of the LSF. %
The region to the right of the line $\theta_1 = 4$ is the domain of
the dominant failure mode $\{g_1(\theta_1,\theta_2) \leq 0\}$ of this
series system. %
The function values $g_2(\theta_1,\theta_2)$ decrease rapidly in
comparison to $g_1(\theta_1,\theta_2)$ in the region near the
origin. %
This leads to an initial steep decreasing gradient of the LSF in the
direction of the secondary failure mode
$\{g_2(\theta_1,\theta_2) \leq 0\}$, as indicated by the contour lines
of the LSF in Fig. \ref{Ex1Fig1}.

\begin{figure}[htb]
  \includegraphics[width=1\textwidth,clip=true,trim= 30mm 37mm 49mm 36mm]{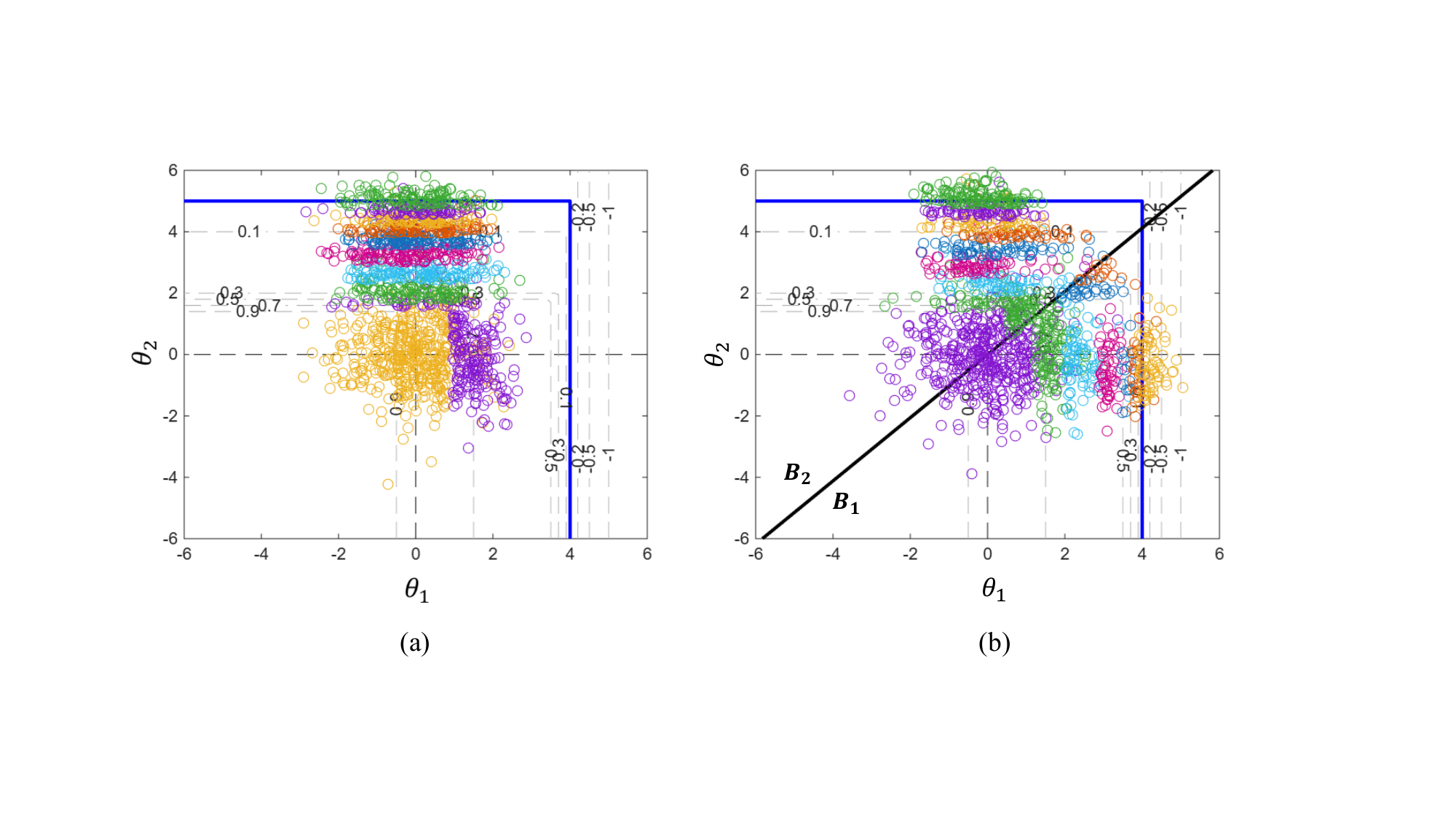}

  \smallbreak

  \includegraphics[width=1\textwidth,clip=true,trim= 30mm 37mm 49mm 36mm]{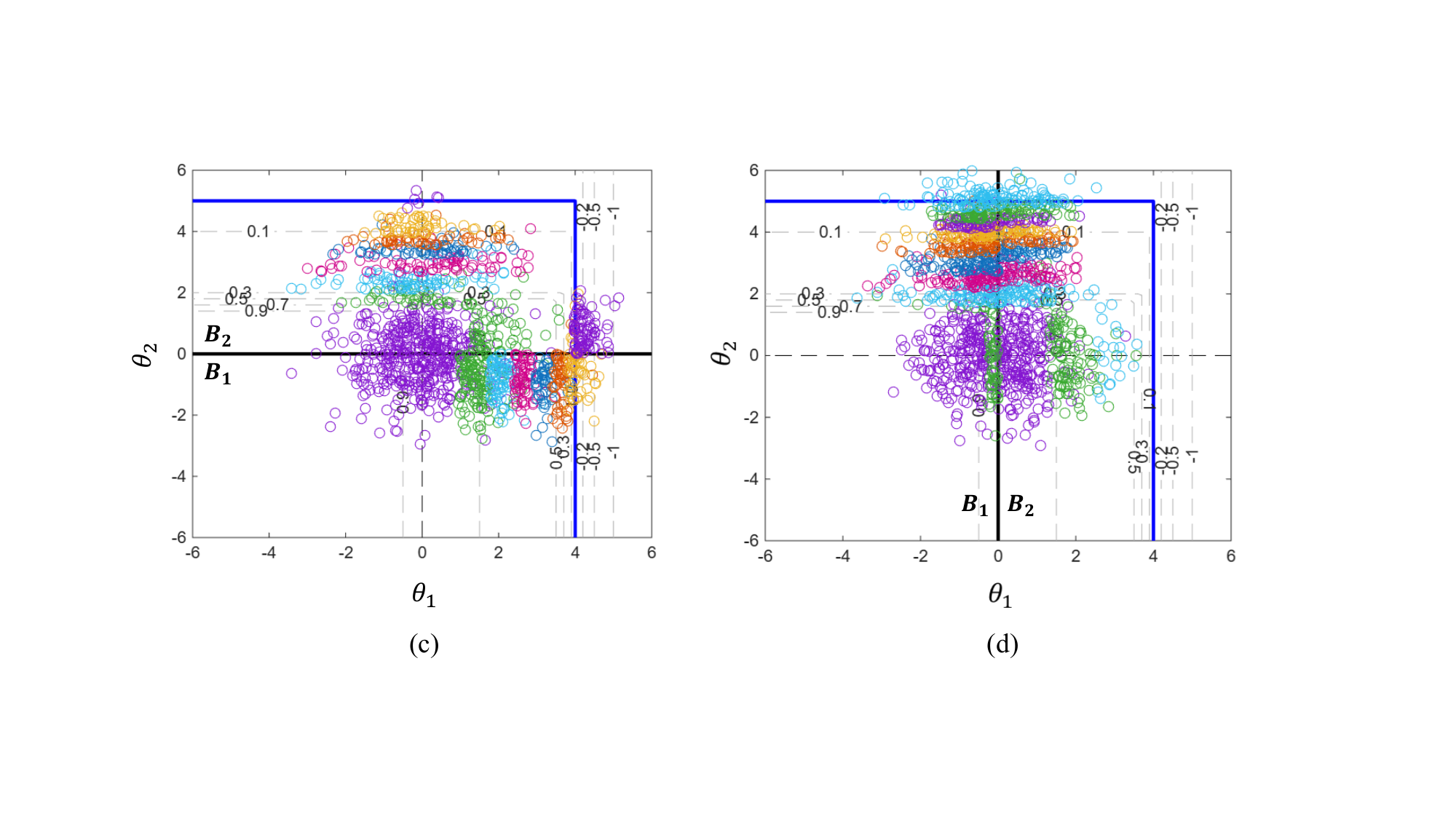}

  \caption{%
    Progression of samples in a typical run of (a) SS and (b--d) Case
    1--3 of the dSS method for the piecewise linear LSF in Example
    4.1, with $N = 500$ samples per level and level probability of
    $\rho = 0.2$. The black line in (b--c) represent the bins used by
    dSS.%
  }%
  \label{Ex4p1_c1_samples}
\end{figure}

To implement the dSS method, we partition the parameter
space into two bins~$B_1$ and $B_2$ of equal probability. We consider three cases:

\begin{enumerate}
\item $B_1 =  \left\{(\theta_1,\theta_2): \mathrm{tan}^{-1}\left(\frac{\theta_2}{\theta_1}\right) \in [-\pi+0.8,0.8]\right\}$ 
and $B_2 = \overline{B_1}$.

\item $B_1 =  \left\{(\theta_1,\theta_2): \theta_2  < 0\right\}$ and $B_2 =  \left\{(\theta_1,\theta_2): \theta_2  > 0\right\}$.

\item $B_1 =  \left\{(\theta_1,\theta_2): \theta_1  < 0\right\}$ and $B_2 =  \left\{(\theta_1,\theta_2): \theta_1  > 0\right\}$.
\end{enumerate}

Figs.~\ref{Ex4p1_c1_samples}a and~\ref{Ex4p1_c1_samples}b, respectively, show the progression of the
samples in the intermediate levels of the standard SS and the proposed
dSS method with binning corresponding to Case 1. %
From the second level onward, all samples in SS move towards the
secondary mode due to the steep decreasing gradient of the LSF in that
direction, and the region on the right towards the dominant mode gets
lost from exploration. %
Consequently, the samples in the final level of SS are spread out only
in the region of the secondary mode at the top, which causes a severe
underestimation of the failure probability. %
In contrast, we see that both failure modes remain populated in all
the intermediate levels of the dSS method, and the samples in the
final level provide a complete representation of the failure domain of
the series system.

\begin{table}[phtb]
  \centering
  \caption{Probability of failure estimates for the piecewise linear
    LSF in Example~\ref{sec:piecewise-linear}, obtained from
      $m = 10^4$ independent runs of each algorithm. The reference value of the probability of failure obtained from
brute-force Monte Carlo simulation with $10^8$ LSF
evaluations is $\hatPFref = 3.19\cdot10^{-5}$.}
  \label{table1}
  \begin{tabular}{p{1.5cm}p{1.7cm}p{1.1cm}p{1cm}p{1cm}p{1.7cm}p{1.1cm}p{1cm}p{1cm}}
    \toprule
    & \multicolumn{4}{c}{SS} & \multicolumn{4}{c}{dSS (Case 1)} 
    \\
    \cmidrule(r){2-5}\cmidrule(r){6-9} 
    & $\widehat{E[\hatPF]}$ & $\hat{\delta}_{\hatPF}$ & $R$ & $N_T$ & $\widehat{E[\hatPF]}$ & $\hat{\delta}_{\hatPF}$ & $R$ & $N_T$
    \\
    \midrule
    \rule{0pt}{1pt}
    $N = 250$ & $3.46\cdot 10^{-5}$ & $4.07$ & $2.14$ & $2386$ & $4.00\cdot 10^{-5}$ & $2.50$ & $0.79$ & $2457$
    \\
    \rule{0pt}{16pt}
    $N = 500$ & $3.23\cdot 10^{-5}$ & $2.90$ & $1.89$ & $4527$ & $3.61\cdot 10^{-5}$ & $1.33$ & $0.50$ & $4816$
    \\
    \rule{0pt}{16pt}
    $N = 1000$ & $3.21\cdot 10^{-5}$ & $1.92$ & $1.56$ & $8428$ & $3.39\cdot 10^{-5}$ & $0.87$ & $0.33$ & $9562$
    \\
    \rule{0pt}{16pt}
    $N = 4000$ & $3.26\cdot 10^{-5}$ & $0.92$ & $0.71$ & $29347$ & $3.25\cdot 10^{-5}$ & $0.40$ & $0.17$ & $37845$
    \\
    \bottomrule
  \end{tabular}

  \vspace*{5mm}
  
  \begin{tabular}{p{1.5cm}p{1.7cm}p{1.1cm}p{1cm}p{1cm}p{1.7cm}p{1.1cm}p{1cm}p{1cm}}
    \toprule
    & \multicolumn{4}{c}{dSS (Case 2)} & \multicolumn{4}{c}{dSS (Case 3)} 
    \\
    \cmidrule(r){2-5}\cmidrule(r){6-9} 
    & $\widehat{E[\hatPF]}$ & $\hat{\delta}_{\hatPF}$ & $R$ & $N_T$ & $\widehat{E[\hatPF]}$ & $\hat{\delta}_{\hatPF}$ & $R$ & $N_T$
    \\
    \midrule
    \rule{0pt}{16pt}
    $N = 250$ & $3.97\cdot 10^{-5}$ & $2.95$ & $0.66$ & $1997$ & $4.14\cdot 10^{-5}$ & $4.38$ & $2.13$ & $2605$
    \\
    \rule{0pt}{16pt}
    $N = 500$ & $3.74\cdot 10^{-5}$ & $2.13$ & $0.51$ & $3843$ & $3.51\cdot 10^{-5}$ & $2.96$ & $1.88$ & $5093$
    \\
    \rule{0pt}{16pt}
    $N = 1000$ & $3.47\cdot 10^{-5}$ & $1.33$ & $0.39$ & $7473$ & $3.41\cdot 10^{-5}$ & $2.08$ & $1.58$ & $10025$
    \\
    \rule{0pt}{16pt}
    $N = 4000$ & $3.29\cdot 10^{-5}$ & $0.65$ & $0.24$ & $28868$ & $3.27\cdot 10^{-5}$ & $0.96$ & $0.75$ & $39787$
    \\
    \bottomrule
  \end{tabular}

  \vspace*{1cm} 

\end{table}

\begin{figure}[phtb]
  \includegraphics[width=1\textwidth,clip=true,trim= 27mm 38mm 31mm 44mm]{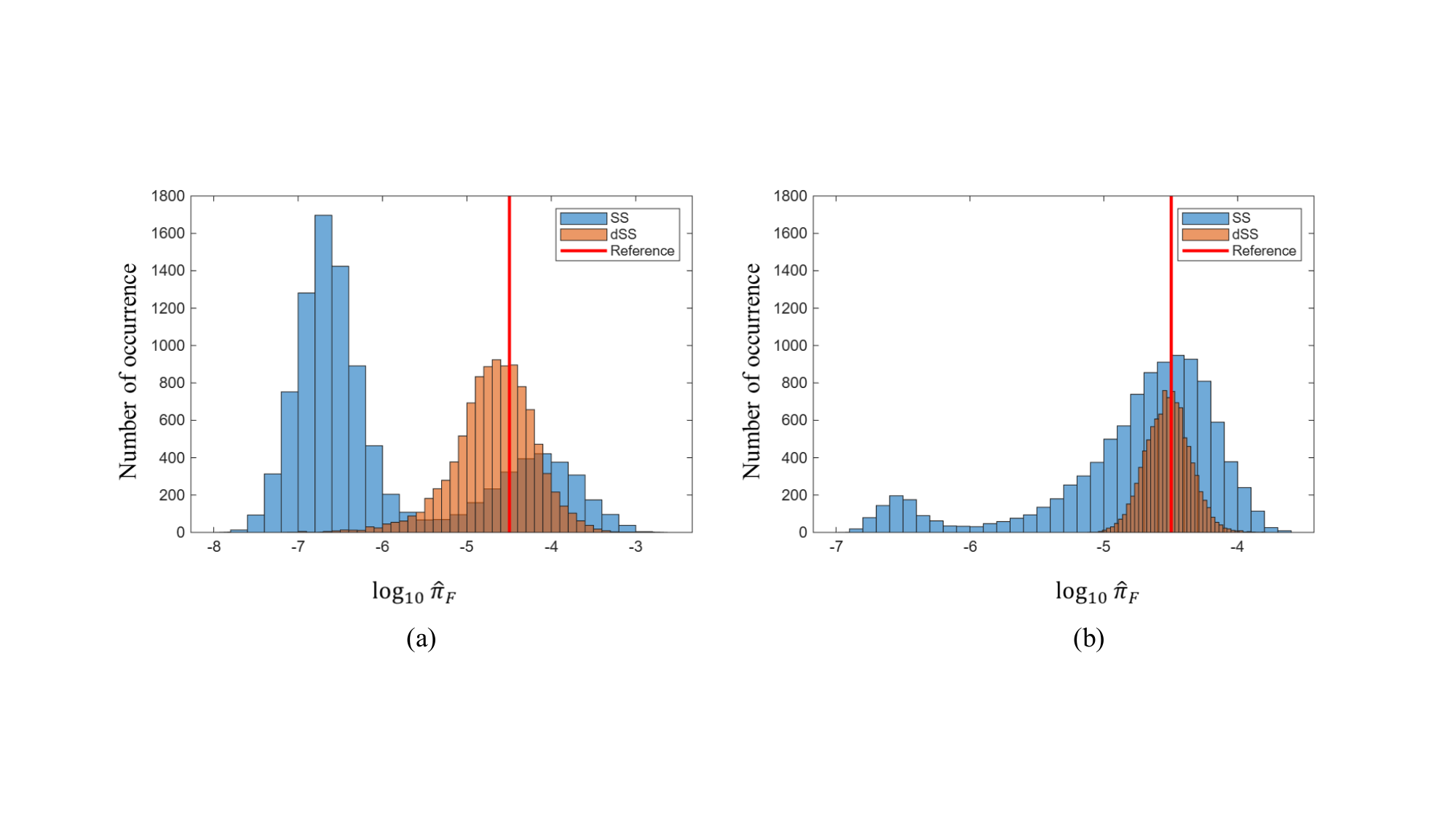}
  \caption{Histogram of the $\log_{10}\hatPF$ estimates for
    the piecewise linear LSF in Example 4.1 obtained from $10^4$
    independent runs of the SS and Case 1 of the dSS methods, with (a) $N = 500$ and (b)
    $N = 4000$ samples per level and level probability of
    $\rho = 0.2$. The red line shows the reference value
    $\log_{10}\hatPFref$.}
  \label{Ex4p1_c1_hist}
\end{figure}

Table~\ref{table1} provides a quantitative comparison of the SS and
the dSS methods in terms of the statistics of the estimates of $\PF$
and the average number of LSF evaluations, denoted by $N_T$. %
The reference value of the probability of failure obtained from
brute-force Monte Carlo simulation with $10^8$ LSF
evaluations is $\hatPFref = 3.19\cdot10^{-5}$. %
$\widehat{E[\hatPF]}$ and $\hat{\delta}_{\hatPF}$, respectively, denote the mean
and the coefficient of variation of the sample estimates of $\PF$ and
the metric~$R$, given by
\begin{equation*}
  R = \sqrt{\frac{1}{m}\, \sum_{i=1}^{m} \left(\log_{10} \left( \frac{\hatPF^{(i)}}{\hatPFref} \right)\right)^2},
\end{equation*}
is the root-mean square error of the failure probability estimates in
the log-scale. %
Fig.~\ref{Ex4p1_c1_hist} shows the histogram of the probability of
failure estimates from $m = 10^4$ independent runs. %
The results indicate superior performance of the dSS method in
comparison to SS in Cases 1 and 2. %
Although the number of LSF evaluations in dSS is marginally higher, SS
grossly underestimates the failure probability in majority of the
runs, as depicted in Fig.~\ref{Ex4p1_c1_hist}. %
In contrast, the estimates from dSS have better concentration around
the reference value. %
Consequently, the estimates from SS have significantly larger
root-mean square error. %
Fig.~\ref{Ex4p1_c1_hist} also depicts a smaller spread in the
estimates from dSS, which leads to a smaller coefficient of
variation. %
A comparison of $\widehat{E[\hatPF]}$ with $\hatPFref$ shows that the
estimates from dSS have a slightly positive bias, which improves and
becomes comparable to standard SS with increase in the per level
sample size $N$. %
The performance of dSS, however, deteriorates in Case 3 and is indeed
similar to SS. %
This is due to the poor choice of partition, where the main failure
mode is located together with half of the smaller one in bin~$B_1$.

\begin{figure}[b]
  \includegraphics[width=1.02\textwidth,clip=true,trim=32mm 35mm 33mm 36mm]{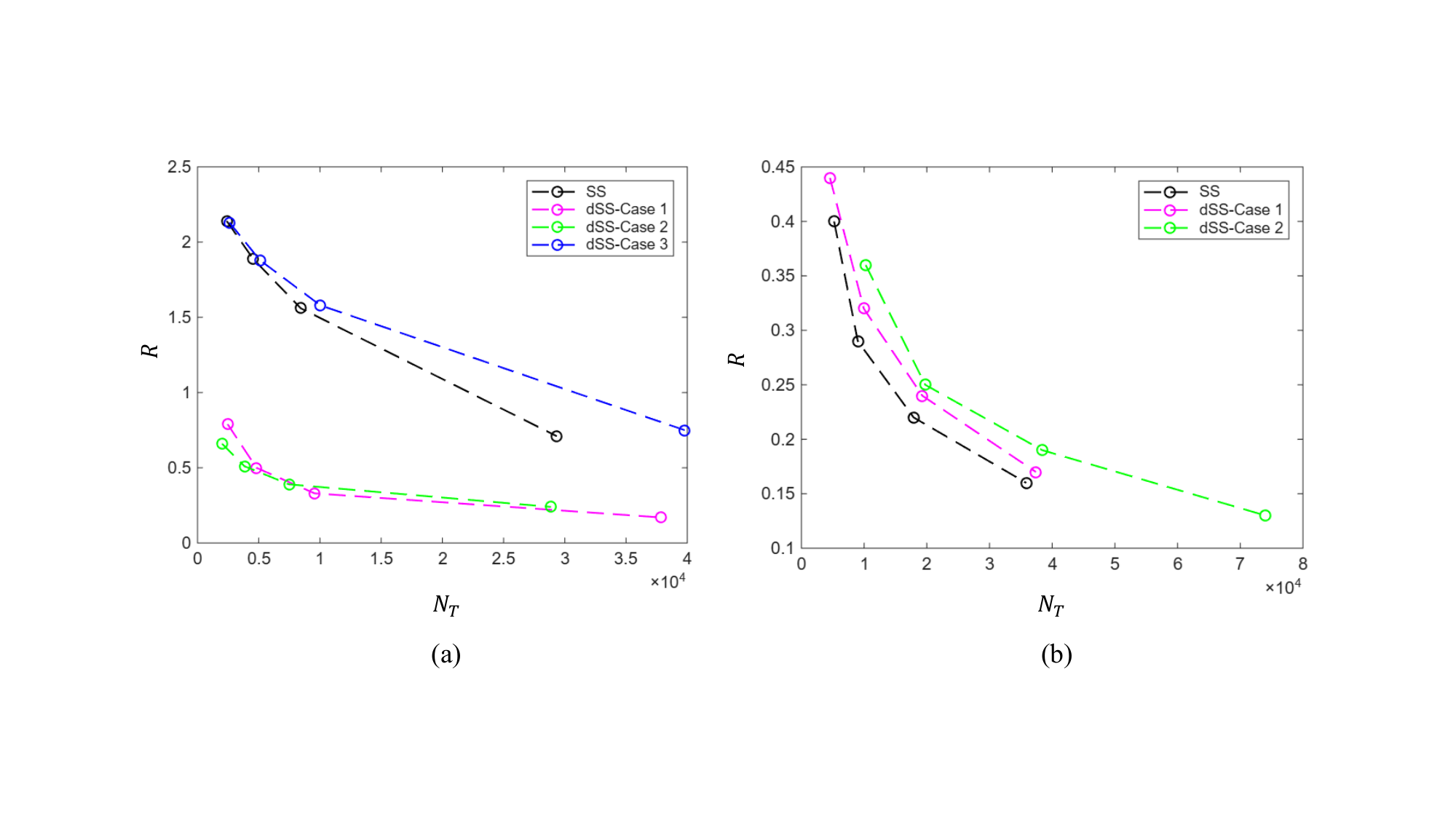}
  \caption{Accuracy of the algorithms, as measured by the metric~$R$,
    as a function of the average number of LSF evaluations~$N_T$. %
    Left: Example~4.1. Right: Example~4.2.%
  }
  \label{RvsNT}
\end{figure}

The performance metric~$R$ is represented as a function of the average
number of LSF evaluations~$N_T$ in Figure~\ref{RvsNT}a.%

\subsection{Several beta points}
\label{sec:several-beta}

\begin{figure}[tp]
  \sidecaption
  \includegraphics[width=0.55\textwidth,clip=true,trim=82mm 36mm 100mm 28mm]{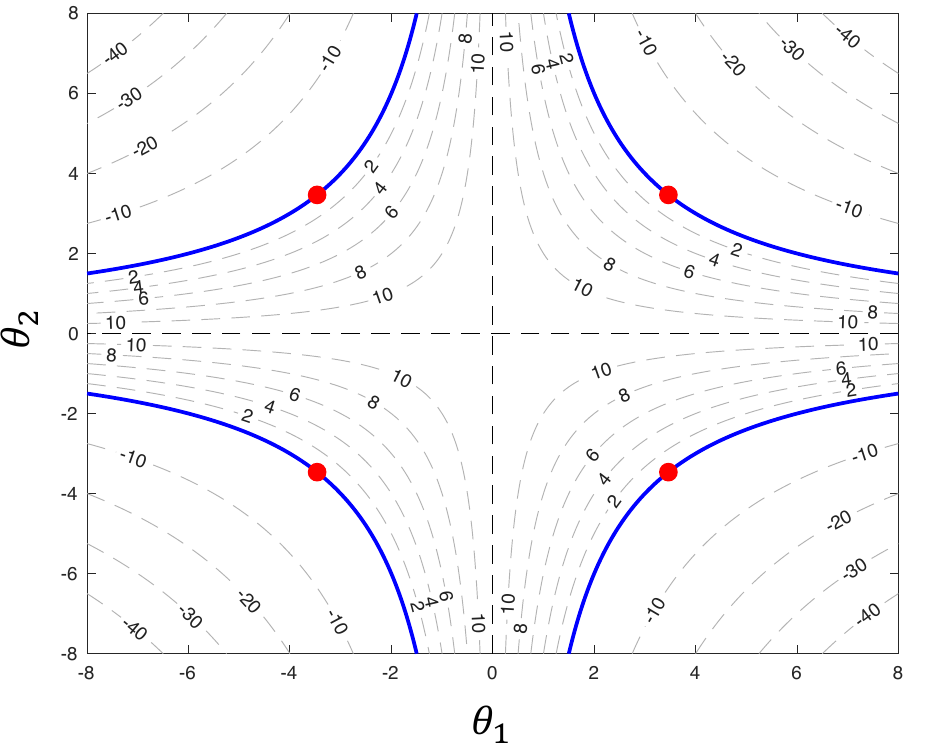}
  \caption{Contour plot of the LSF with several beta points in Example
    4.2. The blue line denotes the boundary of the failure domain. The
    red dots are the four global design point, equidistant from the
    origin $(\theta_1 = 0,\theta_2 = 0)$.}
  \label{Ex2Fig1}
\end{figure}

In this example, we consider the LSF
\begin{equation}
  g(\theta_1,\theta_2) = 12 - \lvert\theta_1\theta_2\rvert, 
\end{equation}
where $\theta_1$ and $\theta_2$ are standard Gaussian random
variables. %
Due to the symmetry of the LSF about the axes $\theta_1 = 0$ and
$\theta_2 = 0$, there are four failure modes with equal importance, as
shown in Fig.~\ref{Ex2Fig1}. %
The reference value of the failure probability obtained from
brute-force Monte Carlo simulation with $10^8$ LSF
evaluations is $\hatPFref = 1.33\cdot10^{-6}$.

\begin{figure}[p]
  \includegraphics[width=1.02\textwidth,clip=true,trim=5mm 25mm 2mm 15mm]{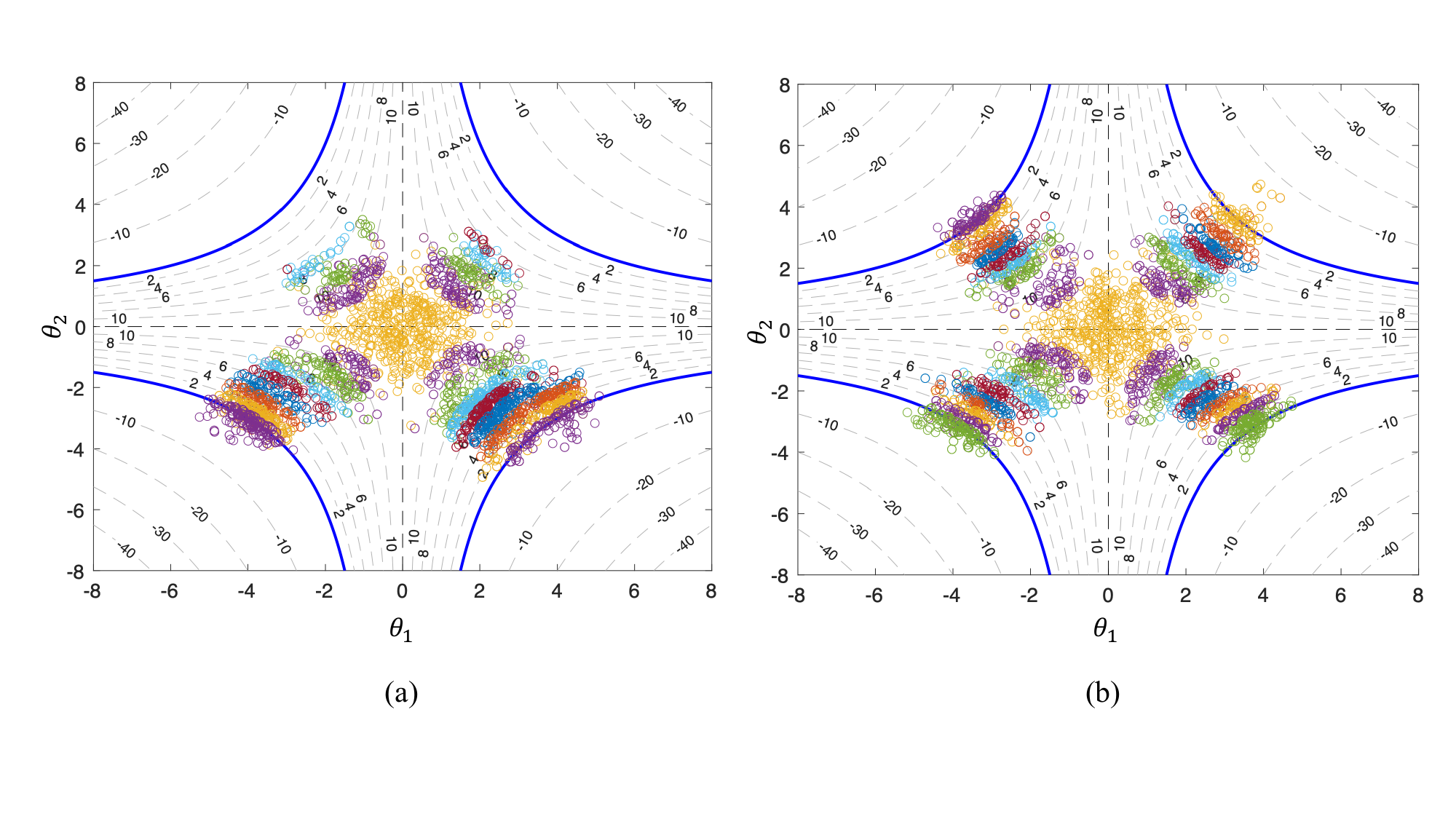}
  \caption{Progression of samples in a typical run of (a) SS and (b) Case 1 of the
    dSS methods for the LSF with several beta points in Example 4.2,
    with $N = 500$ samples per level and level probability of
    $\rho = 0.2$.}
  \label{Ex2Fig2}
\end{figure}

\begin{figure}[tphb]
  \sidecaption
  \includegraphics[width=0.6\textwidth,clip=true,trim=78mm 38mm 115mm 43mm]{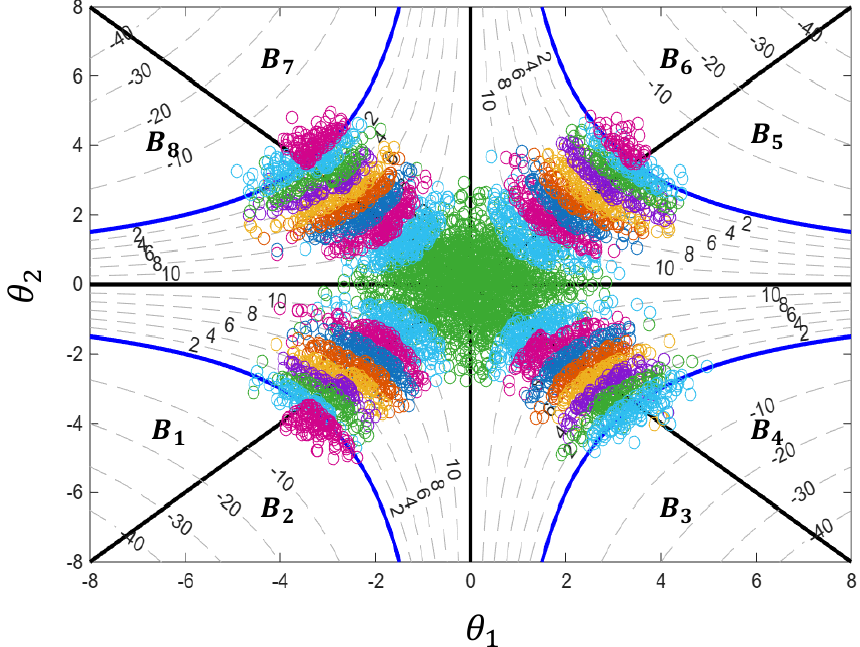}
  \caption{Progression of samples in a typical run of Case 2 of the
    dSS method for the LSF with several beta points in Example 4.2,
    with $N = 2000$ samples per level and level probability of
    $\rho = 0.2$.}
  \label{Ex4p2_c2_samples}
\end{figure}

Figs.~\ref{Ex2Fig2} and~\ref{Ex4p2_c2_samples} show the progression of samples
in a typical run of the dSS and SS methods. %
To implement the dSS method, we consider two choice of partitions.
In Case 1, we partition the parameter space into
four bins of equal probability, which correspond to the four quadrants
in the plane:
\begin{align*}
  B_1 = \left\{ \theta_1 \ge 0 \;\&\; \theta_2 \ge 0 \right\},\quad
  & B_2 = \left\{ \theta_1 \le 0 \;\&\; \theta_2 \ge 0 \right\},\\
  B_3 = \left\{ \theta_1 \le 0 \;\&\; \theta_2 \le 0 \right\},\quad
  & B_4 = \left\{ \theta_1 \ge 0 \;\&\; \theta_2 \le 0 \right\}.
\end{align*}
In Case 2, we consider a partition of eight equal bins, with each bin
representing one half of the four quadrants. %
We see that SS fails to propagate the samples towards all the four
failure modes, but with dSS all modes are explored. %

A quantitative comparison of the performance of the two methods based
on $10^4$ independent runs is reported in Table \ref{table_Ex4p2_1}. %
In contrast to Example 4.1, inadequacy in exploring the entire failure
domain does not cause SS to underestimate the failure probability,
because all the failure modes in this example have equal importance. %
The superior performance of dSS in exploring all the failure modes
comes at the expense of reduced performance in estimating the global
failure probability as indicated by comparatively higher values of 
$\hat{\delta}_{\hatPF}$ and $R$, particularly in Case 2. %
This is due to the requirement to estimate multiple local thresholds, 
four in Case 1 and eight in Case 2, with the same per level sample size $N$. %
However, this could be a reasonable trade-off in situations where it is important
to be able to identify all the failure modes, and to assess the contribution of 
the individual modes to the global failure probability. %

The performance metric~$R$ is represented as a function of the average
number of LSF evaluations~$N_T$ in Figure~\ref{RvsNT}.%

\begin{table}[p]
  \centering
  \caption{%
    Probability of failure estimates for the LSF with several beta
    points in Example~\ref{sec:several-beta}, obtained from
      $m = 10^4$ independent runs of each algorithm. 
      The reference value of the failure probability obtained from
brute-force Monte Carlo simulation with $10^8$ LSF
evaluations is $\hatPFref = 1.33\cdot10^{-6}$.%
  }%
  \label{table_Ex4p2_1}       
  \begin{tabular}{p{1.5cm}p{1.7cm}p{1.1cm}p{1cm}p{1cm}p{1.7cm}p{1.1cm}p{1cm}p{1cm}}
    \toprule
    & \multicolumn{4}{c}{SS} & \multicolumn{4}{c}{dSS (Case 1)} 
    \\
    \cmidrule(r){2-5}\cmidrule(r){6-9} 
    & $\widehat{E[\hatPF]}$ & $\hat{\delta}_{\hatPF}$ & $R$ & $N_T$ & $\widehat{E[\hatPF]}$ & $\hat{\delta}_{\hatPF}$ & $R$ & $N_T$
    \\
    \midrule
    \rule{0pt}{1pt}
    \rule{0pt}{16pt}
    $N = 500$ &  $1.52\cdot 10^{-6}$ & $2.76$ & $0.40$ & $5213$ & $2.12\cdot 10^{-6}$ & $3.97$ & $0.44$ & $4552$
    \\
     \rule{0pt}{16pt}
    $N = 1000$ & $1.44\cdot 10^{-6}$ & $1.02$ & $0.29$ & $8996$ & $1.69\cdot 10^{-6}$ & $1.45$ & $0.32$ & $9944$
    \\
    \rule{0pt}{16pt}
    $N = 2000$ & $1.41\cdot 10^{-6}$ & $0.86$ & $0.22$ & $17903$ & $1.56\cdot 10^{-6}$ & $0.91$ & $0.24$ & $19280$
    \\
    \rule{0pt}{16pt}
    $N = 4000$ & $1.38\cdot 10^{-6}$ & $0.49$ & $0.16$ & $35855$ & $1.46\cdot 10^{-6}$ & $0.59$ & $0.17$ & $37394$
    \\
    \bottomrule
  \end{tabular}

  \vspace*{5mm}
  
  \begin{tabular}{p{1.5cm}p{1.7cm}p{1.1cm}p{1cm}p{1cm}}
    \toprule
    & \multicolumn{4}{c}{dSS (Case 2)}
    \\
    \cmidrule(r){2-5}
    & $\widehat{E[\hatPF]}$ & $\hat{\delta}_{\hatPF}$ & $R$ & $N_T$
    \\
    \midrule
    \rule{0pt}{1pt}
    $N = 1000$ &  $2.18\cdot 10^{-6}$ & $2.47$ & $0.36$ & $10249$
    \\
    \rule{0pt}{16pt}
    $N = 2000$ &  $1.73\cdot 10^{-6}$ & $1.36$ & $0.25$ & $19749$
    \\
    \rule{0pt}{16pt}
    $N = 4000$ & $1.56\cdot 10^{-6}$ & $0.73$ & $0.19$ & $38411$
    \\
    \rule{0pt}{16pt}
    $N = 8000$ & $1.46\cdot 10^{-6}$ & $0.39$ & $0.13$ & $74025$
    \\
    \bottomrule
  \end{tabular}

  \vspace*{1cm} 
\end{table}

\section{Conclusions}
\label{sec:conclusions}

We propose a new advanced Monte Carlo method called directional subset
simulation (dSS) to estimate small failure probabilities. %
dSS is an enhancement of the standard SS method for reliability
analysis. %
It retains the generic multi-level nature of SS, wherein the small
failure probability is estimated as a product of larger conditional
probabilities of a nested sequence of intermediate failure events, or
subsets, defined over the parameter space. %
However, the construction of the subsets in dSS differs from that in SS.

dSS works by partitioning the parameter space into non-overlapping
bins. %
At each level, the global subset over the parameter space is a union
of local subsets within each bin, %
defined in terms of a local threshold parameter that
is estimated from the LSF values of the samples in the bin. %
As in SS, global MCMC moves are used to propagate samples from one
level to the next. %
Numerical examples demonstrate that the use of multiple threshold
parameters at each level provides an effective means to account for
sudden sharp changes in the gradient of the LSF in different
directions, which is an advantage over standard SS. %
It also facilitates adequate exploration of failure domains with
multiple modes lying in different directions over the parameter
space. %

The choice of bins influences the performance of the dSS method. %
In many real-life engineering systems, experts are likely to have a
prior idea of the failure modes of the system. %
This knowledge, when it is available, can be used to select the
bins. %
When such knowledge is not available, the situation is more
difficult. %
The naive ``default'' choice of using orthants results in very large
number of bins if the input dimension large, %
which is a problem since each bin has its own local threshold that
must be estimated.%

A possible approach to address this issue is to develop an adaptive mechanism
that automatically creates the bins as the algorithm progresses. %
One possibility is to start with a single run of standard SS,
i.e., a single bin encompassing the parameter space, and use a suitable
diagnostic measure to detect multimodality in the failure domain. %
In this context, unsupervised learning techniques such as clustering based
on the LSF values of the samples in each conditional level can be applied
to detect the important directions containing the failure modes, and accordingly 
split the parameter space to create the bins. %
Future research will focus on incorporating the above ideas into the current 
framework of the dSS method. %
Another direction for future work is to compare the dSS method
with other variants of SS, such as
\citep{Sharma_PEM_2023, Rakshi_MSSP_2021, Kinnear_PEM_2025}.

\section*{Acknowledgment}

The author O. Kanjilal acknowledges the German Academic Exchange
Service (DAAD) for support through the Short-term Postdoc Grant
(No. 57611805), and the Nemetschek Innovation Foundation for support
through the TUM-GNI Postdoc Program.

\bibliographystyle{plainnat}
\bibliography{Kanjilal_Bect_ENUMATH}

\end{document}